\documentclass[letterpaper,twocolumn,10pt]{article}

\usepackage{usenix2020_SOUPS}
\usepackage[nobiblatex]{xurl}

\usepackage{tikz}
\usepackage{amsmath}

\usepackage{amssymb,graphicx,color, tikz, pifont, makecell}
\usepackage{amsfonts, enumerate, booktabs}
\usepackage{latexsym}
\usepackage{amsmath}
\usepackage{setspace}
\usepackage{float, threeparttable}
\usepackage[toc,page]{appendix}
\usepackage{pdfpages}
\usepackage{enumitem}
\usepackage{multirow}
\usepackage{svg}
\usepackage{pgf-umlsd}
\usepackage{wasysym}
\usepackage{pdfpages}
\usepackage{placeins}
\usepackage{flowchart}
\usepackage{pgf-pie, pgfplots, pgfkeys}
\usepackage{pgfplotstable}
\pgfplotsset{compat=1.7}
\usetikzlibrary{shapes,arrows, arrows.meta,chains,external, positioning}

\definecolor{checkgreen}{rgb}{0.09, 0.45, 0.11}
\definecolor{crossred}{rgb}{0.72, 0.05, 0.04}
\definecolor{maybeyellow}{rgb}{1, 0.80, 0.04}

\newcommand{\up}[1]{\color{#1}\UParrow\color{black}}%
\newcommand{\down}[1]{\color{#1}\DOWNarrow\color{black}}%
\usetikzlibrary{matrix,calc}

\newcolumntype{L}[1]{>{\hsize=#1\hsize\RaggedRight} X}
\newcommand{\eat}[1]{}

\newcommand{\tss}{
    \tikzset{>={Latex[width=2mm,length=2mm]}}
    \tikzstyle{line} = [draw, ->, >=latex, ultra thick]
    \tikzstyle{circ} = [
        circle,
        align=center,
        text width=2em,
        text centered,
        inner sep=1mm,
        outer sep=1mm,
        minimum width=0cm,
        minimum height=0cm
    ]
    \tikzstyle{tc} = [
      circle,
      draw,
      ultra thick,
      align=center,
      text width=2em,
      text centered,
      inner sep=1mm,
      outer sep=1mm,
      minimum width=0cm,
      minimum height=0cm
    ]
    \tikzstyle{csm} = [
        circle,
        align=center,
        text width=2em,
        text centered,
        inner sep=0mm,
        outer sep=0mm,
        minimum width=0cm,
        minimum height=0cm
    ]
    \tikzstyle{block} = [
        rectangle,
        draw,
        text centered,
        align=center,
        minimum width=10em,
        minimum height=10em
    ]
    \tikzstyle{noshape} = [text width=5em, text centered, minimum height=5em]
}
\usetikzlibrary{matrix,calc}

\begin{document}

\date{}

 
\title{\Large \bf A Decentralised Digital Token Architecture for Public Transport}

\def\plainauthor{Author name(s) for PDF metadata. Don't forget to anonymize for submission!}

\author{
{\rm Oscar King}\\
University College London
\and
{\rm Geoffrey Goodell}\\
University College London
} 

\maketitle

\begin{abstract}
Digitisation is often viewed as beneficial to a user. Whereas traditionally, people would be required to physically identify to a service, pay for a ticket in cash, or visit a library to access a book, people can now achieve all of this through a click of a button. Such actions may seem functionally identical to their analogue counterparts, but in the digital case, a user's actions are automatically recorded. The recording of user's interactions presents a problem because once the information is collected, it is outside of the control of the person whom it concerns. This issue is only exacerbated by the centralisation of the authentication mechanisms underpinning the aforementioned services, permitting the aggregation and analysis of even more data. This work aims to motivate the need and establish the feasibility of the application of a privacy-enhancing digital token management service to public transit. A proof-of-concept implementation is developed, building upon a design proposed by Goodell and Aste. This implementation was optimised for the public transport use case. Its performance is tested in a local environment to better understand the technical challenges and assess the technical feasibility of the system in a production setting. It was observed that for loads between one and five requests per second the proof-of-concept performs comparably to other contactless payment systems, with a maximum median response time less than two seconds.  Due to hardware bottlenecks, reliable throughput in our test environment was limited to five requests per second. The demonstrated throughput and latency indicate that the system can feasibly compete with solutions currently in use. Yet, further work is needed to demonstrate their performance characteristics in an environment similar to that experienced in production.
\end{abstract}

\section{Introduction}\label{intro}
Every day, millions of users around the world interact with digitalised public services. User authentication is often required to gain access to the service. Users likely trust that during this authentication process their personal information is processed privately and securely.  Also, users seek interfaces that are swift, well-planned, and clear; they can quickly become frustrated or distrust services that are difficult to use.  Ultimately, they may avoid a service entirely if the usability poses a significant obstacle to achieving their primary goal~\cite{Brostoff}.

\subsection{Problem Statement}
Current widely-used Identity Management Systems (IDMS) such as \textit{GOV.UK Verify} and the \textit{Federal Cloud Credential Exchange} (FCCX) have demonstrated shortcomings related to privacy and security~\cite{Brandao}. These concerns are exacerbated by the following two factors: 
\begin{enumerate}
    \item The burgeoning growth of systems, both in terms of services and in terms of users~\cite{ElMaliki2007}.
    \item Such systems adopting federated or centralised authentication systems. The introduction of such a central hub makes it possible to aggregate and link a user's interactions with different service providers, as well as to have broad access to personally identifiable information (PII). 
\end{enumerate}
More users or more linked services make a target more valuable for malicious attackers or malicious hub operators, and in the case of compromise by these actors, it becomes possible to undetectably impersonate, profile, and target individual users~\cite{Brandao}, leading to an erosion of public trust~\cite{Ohara} in the service. Unfortunately, issues with privacy and security are not limited to government-run services. Recently reported vulnerabilities or data breaches in platform services operated by Facebook and Google, which are often used for user authentication to third-party services, are important examples~\cite{fbreach, googlebreach}.

Responding to these issues with the logic that the best way to uphold trust is through privacy by design, research has been undertaken in privacy-enhancing identity and credential management systems~\cite{Goodell2019, Idemix, UPort, Chainspace, AnonCredLight}.  Individually, these systems address a subset of privacy and security concerns, each with their own merits, privacy properties, trade-offs, and drawbacks~\cite{Goodell2019}. However, the interplay between the specific merits and limitations of these forms of systems in the context of concrete use cases are not always well-understood.  Different use-cases have different design requirements due to the specific needs of the system's stakeholders.  Careful analysis is therefore required to determine the specific considerations system implementers and users must consider, understand, and address.  Further study is also required to assess the extent to which current privacy-enhancing identity and credential management systems can feasibly compete with systems currently deployed and used by enterprises and governments.

\subsection{Research Objectives}
This chapter investigates the feasibility and effectiveness of implementing privacy-enhancing identity management in large-scale public transport payment systems. To limit the scope of the work we explore the following three questions in the context of the protocol described by Goodell et al~\cite{Goodell2019}, to which we refer as the \textit{Distributed Digital Identity Architecture} (DigID):

\begin{enumerate}
    \item How does the DigID address the key privacy and security requirements of modern public transport payment systems compared to current centralized solutions?
    
    \item What are the technical, operational, and user experience trade-offs when implementing DigID in a large-scale public transport network?
    
    \item To what extent can DigID meet or exceed the performance requirements of Transport for London's existing payment infrastructure?
\end{enumerate}

We use London as a case-study due to it's scale, fare structure, and readily available public data against which we can compare results.

\subsection{Chapter Structure}
The chapter continues as follows. The next section provides context for such a system and helps motivates their utility. Additionally, it introduces terminology and possible competing solutions. Section~\ref{overview} describes the challenges in applying our proposed solution to Transport for London (TfL) and suggests possible extensions to address the limitations.  Section~\ref{implementation} details the design specific choices in the implementation.  Section~\ref{results} describes the testing approach, Section~\ref{analysis} discusses the results, and Section~\ref{conclusion} offers an analysis, with a view toward future work.

\section{Background}\label{background}
The regulatory and economic context is constantly evolving with the rise of the digital economy and new regulations facing financial institutions, particularly with respect to customer verification and with respect to non-fiat digital currencies.  In this section, we characterise that context as well as the key concepts underpinning this project and important related work.
\subsection{Societal Context}

Public transportation remains a vital service for millions of people worldwide. In London, TfL reports that, on average, over nine million trips were taken per day on the network in its fiscal year ending 2023~\cite{tfl2023}, not including non-TfL rail journeys within London, such as those provided by the National Rail network. By comparison, in New York City, the Metropolitan Transportation Authority (MTA) reports that about five million trips are taken each day on its subway and bus network on an average day~\cite{nyc2023}, not including MTA commuter rail journeys, or rail and bus journeys in and around New York City provided by other public transportation authorities, such as New Jersey Transit and Port Authority Trans-Hudson. Although some of these trips might not constitute essential travel, many millions rely on both transport networks for their daily commutes into work.

There exist many reasons a commuter might choose to take public transport over other modes of travel. For some public transport may be quicker, more convenient, or more amenable to multi-tasking. Some commuters may not be physically able to walk, cycle, or drive a car.  Additionally, some may simply not have any other option. For any of these commuters, but especially ones where an alternative mode does not readily exist, access to private, reliable, public transportation is indispensable.

With the rise of the digital economy, the world has seen more collection, aggregation, and analysis of personal data than ever before, undermining the foundational privacy of individuals and households.  There is no question that data flows are tremendously valuable to both public-sector and private-sector actors.  Because personal data can always be used to discriminate and exclude individual persons on the basis of their profiles, it is not possible to guarantee that public transportation remains truly open to the public unless their privacy is protected.  Data protection, or ``privacy by promise,'' is insufficient to meet this requirement, since promises can always be broken: There are countless examples of consumers being profiled as the result of data breaches by government requests, unscrupulous insiders, and a plethora of attackers ranging from small-scale criminals to well-funded nation states.

If one is not convinced by the axiom that privacy is a fundamental human right, as enshrined by the Universal Declaration of Human Rights (Article 12), the European Convention of Human Rights (Article 8), and the European Charter of Fundamental Rights (Article 7), then other rationales are worth considering as well.  The design of payment mechanisms supporting many modern transport systems, including, among others, \textit{Oyster} in London and \textit{OMNY} in New York, encourages individual passengers to pay with bank accounts, which are by law linked to their identities.  Even when not directly linked to personal identifiable information, repeated use of the same access token (for example, a travel card) over time creates profiles of users that can then be analysed and combined with other information to be linked to unique individuals.  The use of credit, for example, by allowing individuals to provide an identity and subsequently pay after travelling, introduces a secondary issue of how debts are handled.  In all cases, the creation of profiles not only impinges upon the privacy of individuals but also introduces the means to exclude or restrict particular users.  There exist varying levels of trust in government jurisdictions around the world~\cite{oecd2024}, and although a given government of a particular jurisdiction might lack the appetite to implement this kind of surveillance, there is no way to be sure that future governments with access to the same infrastructural capabilities will be similarly conscientious.  One oft-cited example of how surveillance of this kind can lead to exclusion from public services is the Social Credit System in China, wherein people have been excluded from accessing trains as well other services on the basis of their past behaviours and other factors~\cite{mcdonald2019}.

Over the past decade, large-scale publicly funded digital services offered by transportation networks have undergone significant modernisation.  Various examples of modernisations include:

\begin{itemize}

\item contactless payments in London in 2014~\cite{Tfl-Contactless},

\item the roll-out of OMNY by the Metropolitan Transportation Authority in New York in 2019~\cite{OMNY, NYC-Contactless}, and

\item the projected adoption of electronic payment methods for public transit in the Parisian metropolitan area~\cite{Paris-Contactless}.

\end{itemize}

It is argued that large institutions are capable of implementing change only when afforded the right incentives; otherwise, the above-mentioned modernisation might not have taken place. In the case of contactless payments for public transport, modernisation makes the service easier to use and likely increases usage~\cite{Lambrecht2012}. Where transit is government-run, profit maximisation is not the main goal. Instead, these service operators are typically looking to increase consumer surplus and social good. Therefore, increased public usage can be leveraged to decrease price, as well as to generate additional revenue to allow further investment to improve services. By extension of this argument, the introduction of contactless payment by these institutions should aid in the adoption and implementation of payment protocols that support privacy by design, given the right economic incentives.  At least two economic justifications support the development of such protocols:

\begin{itemize}

\item\textit{Changing public attitude.} People are becoming more aware of the data that are being collected and the value of such data~\cite{POLITICO2019, Akamai}. Consumers in the European Union are even willing to forego savings on products and services to preserve rights afforded to them under the General Data Protection Regulation (GDPR)~\cite{Godel2017}. This willingness to forego savings supports the claim that data privacy can be a significant hurdle when deciding whether to use a service. It is shown that reducing these hassle costs\footnote{Not to be confused with inconvenience.} can increase revenue and user retention~\cite{Lambrecht2012}.

\item\textit{Potential cost savings.}  First, depending on the use case, the costs and risk of maintaining specialised infrastructure can be high. Service providers may wish to outsource the management of system sub-components that are not part of their core businesses. Service provides have previously taken these steps, such as the operation of OMNY and TfL by third parties~\cite{OMNY-Cubic, TfL-Cubic}. Second, the operational costs of the protocol could be much lower than current alternatives. In 2016, TfL awarded Barclaycard a ten-year contract worth up to £380 million as its merchant acquirer~\cite{TransportForLondon-Contract}. A cost model developed by Ernst and Young~\cite{Cost-of-blockchains} suggests that a similar task can likely be completed with substantial savings.

\end{itemize}

As the public attitudes towards personal data and data privacy have evolved, regulations concerning personal data and data privacy have become more stringent~\cite{GDPR}. The European Union's General Data Protection Regulation (GDPR) stipulates how, under what circumstances, and for what reasons personal data can be processed, stored, and managed, including provision for the levying of substantive fines against firms or institutions that are found to be non-compliant~\cite{BA-fine, google-fine}.  The GDPR has also inspired the adoption of similar regulations in jurisdictions outside Europe.

Systemically limiting the amount of personal data that a system can collect by virtue of its design may be a worthwhile endeavour for the system operators.  Fundamentally distinct from designing a system that does not hold data following use, an approach based on privacy by design can limit the operator's exposure to security and compliance risks of associated with collecting personal data.  Mitigating these risks may result in the decreased likelihood of fines as well as the possibility of improved public image and trust. In the case of a third party platform services provider without public funding, the decreased risk of fines is likely the stronger motivator, because fines directly impact companies' profits.  For a public institution, the increased trust and improved public image are likely to be of primary concern.  Compared with a system that is private by design, a system that is merely ``private by promise'' does not enjoy these benefits, as it relies on unverifiable trust that data are indeed deleted or used only for the specified purposes, and therefore there is always an inherent risk of data misuse.

\subsection{Distributed Ledger Technology and Associated Terminology}

In the literature, the term ``blockchain'' is occasionally used interchangeably with ``distributed ledger technology (DLT) system'', although they are however distinct concepts. Distributed ledger technology refers to the network of distinct participants, the storage mechanism, and the consensus algorithm, a blockchain is the underlying data-structure with which many DLT systems are built.

Distributed ledgers can take different forms and support many purposes and use cases, each requiring a different kind of access control. The terminology pertaining to ledger access control and use is sometimes inconsistent throughout literature; we use terminology consistent with ISO 22739:2024~\cite{iso22739}. There are two dimensions along which DLT systems are typically classified in the context of access control. Along one axis a distinction is made between \textit{public} and \textit{private} DLT systems. Along the other the distinction is made between \textit{permissioned} and \textit{permissionless} DLT systems. The terms \textit{public} and \textit{private} refer to the use of the DLT system and the benefit derived from its use. \textit{Public} ledgers allow anyone to benefit either directly or indirectly from its use. \textit{Private} ledgers only provide utility for a fixed set of participants. The terms \textit{permissioned} and \textit{permissionless} refer to the manner by which entities are allowed to access the DLT to either write to it or read from it. \textit{Permissioned} DLT systems restrict access to an authenticated group of participants. \textit{Permissionless} DLT systems do not restrict access in this manner and are open to full participation and validation by anyone who wishes.

The combination of these configurations may provide desirable network properties. Public-permissioned DLT systems may be used by governmental agencies, so that citizens and businesses can access information but cannot alter this ledger's state.  Private-permissioned DLT systems provide a way for one or more mutually distrusting participants to share an evolving record of history in support of some common goal, e.g. exchanging assets, incrementally developing a document, or performing services~\cite{Androulaki2018}. The protocol described by Goodell et al.~\cite{Goodell2019} makes use of a DLT system that, in principle, can be used by anyone, although it is presumed to be operated by registered and regulated entities. Thus, the scope of our discussion is limited to public-permissioned DLT systems.

\subsection{Privacy-Enhancing Identity Management Systems}\label{privacy-idms}
There are many different possible implementations of a privacy-enhancing IDMS, each with its own set of privacy characteristics. First, to minimise the set of parties that users must trust with their personal data, the protocol must minimise the set of control points that can be used to compromise the system.  Second, to protect the right to privacy and ensure that users have reason to feel free and behave naturally, which in turn could improve adoption of the system, the protocol must eliminate vectors that could enable mass (``dragnet'') surveillance.  Finally, to support personal choice, the system must allow users to manage their own data linkages within the system.  Without these three characteristics, institutions might still be able to profile people based on their actions within the system.  We based our design upon the DigID protocol developed by Goodell and Aste~\cite{Goodell2019}. We make no claims that the characteristics are not also met by other designs.

\subsection{Related work}

Our treatment of this design space builds upon earlier work, beginning with Heydt-Benjamin et al~\cite{privacy-for-public-transportation}, who proposed an early design and argued in 2006 that the problem of ensuring that public transport users would not be profiled on the basis of their journeys is worthy of serious consideration.  We also specifically note a serious treatment of this topic in 2019 by Japp-Henk Hoepman~\cite{hoepman2019} that proposes three approaches to addressing this problem, one with overall poor privacy properties because information-revealing interbank payments must be used for each trip; one that is relatively good in terms of privacy but that relies upon accounts that can be used to link successive journeys to each other; and a solution that has no real privacy because of blacklistable credentials.  In this chapter, we propose a solution that is similar to the second but with tokens rather than balances, and that is more robust than all of them in terms of privacy.

Hoepman considers several primitives that can be used in such a system.  Partially blind signatures are equivalent to fully blind signatures using attribute-specific keys.  However, if there are too many attributes to make multiple keys practical, then, ipso facto, the anonymity set is too small.  Attribute-backed credentials (ABCs) of the type proposed by Camenisch and Lysyanskaya~\cite{camenisch2001} depend upon a root identity, which can be revealed via blackmail, or even by demanding the prover to reveal evidence that the same master key is associated with both a credential in question and a valid credential of which a person can have at most one, such as a taxpayer ID in a particular jurisdiction.  Even if the latter credential is not revealed explicitly, the singularity of the set of attributes is established, which prevents a user from establishing multiple independent identities.  Finally, it must be acknowledged that identity proofing (or insisting upon non-transferability of credentials) is potentially dangerous: Binding a person to a singular device is not the way forward if privacy is the goal.

\section{Adapting the Architecture for Transport for London}\label{overview}
Transport for London (TfL) is the governing body regulating all public transport in the London Metropolitan Area (LMA) apart from National Rail services~\cite{About-TfL}; it governs one of the world's most extensive public transit networks~\cite{Busiest-Metros}. It owns several subsidiaries that control or operate the public transit services provided in the LMA, such as bus and tram services. Henceforth, we refer to TfL in the general sense, to include it and any of its subsidiaries.

Although TfL operates the London Underground system, many other services supporting the network, such as the sale of Oyster credit and the operation of London buses, are performed under contract by independent third parties~\cite{tfl-buses-private}. Much of the operational infrastructure required to support system operators is well-established, as the system is already partially decentralised.  However, although it might be expected that many of the roles required for operating a decentralised payment infrastructure are already clearly defined, this is not the case.  For example, Oyster credit is verified centrally, by a firm called Cubic Transportation Services~\cite{cubic2014}.

The challenges associated with mapping potential participants in a decentralised payment infrastructure onto existing and potential stakeholders, as well as the various procedural and legislative adaptations that might be required for implementation, are outside the scope of this work. Rather, we focus on the extensions and changes required to adapt the existing protocol so that users can interact with a new decentralised payment system in a manner similar to how they interact with the current TfL system and its constraints. Unless explicitly stated otherwise, we assume that the proposed system would retain most of the user interface characteristics of the existing system, including, for example, the use of fare zones and entry gates and exit gates for all trips for which the total fare is not known at the point of entry.  Deviating too much from how users commonly interact with the system might reduce the adoption of the system, by introducing additional barriers.

In the simplest problem setting, the user will interact with the system at three points: (a) to purchase Oyster credit, (b) to gain entry to the system after proving sufficient credit, and (c) if applicable, to exit of the system upon termination of the journey.  This problem setting describes what is commonly referred to by TfL as ``Pay-as-you-go'' (PAYG).  It is important to note that both Oyster and the approach we propose are not general-purpose payment schemes, but actually e-money: tokens represent credit for value that the passenger has already paid through an exogenous medium.

There are two main differences between the way a user would interact with the existing PAYG fare system offered by TfL and the DigID system.  First, PAYG calculates journey cost based on entry and exit stations. The DigID protocol only models a single interaction with an authenticating party (AP) and a service. There is no way for the protocol to know \textit{a priori} where a user entering the system will go. Similarly, it is impossible for the protocol to know \textit{ex ante} where a user exiting the system had originally entered.  Second, when using an Oyster card or payment card with PAYG, devices held by users are not required to store or process volatile data that persist between interactions, whereas with DigID, user devices are required to retrieve and store tokens for later use.  To avoid significantly altering how passengers would interact with the system, these foundational differences must be addressed carefully.  The protocol that we propose adapts the DigID protocol to enable stateful interactions involving user devices without introducing fundamentally new patterns of interaction between the user and the payment system.

\subsection{Protocol Explanation}
\label{explanation}
\newcommand{\figsize}{1}
\newcommand{\RLPos}{(7,-3)}
\newcommand{\RHPos}{(7,3)}
\newcommand{\APos}{(-7,0)}
\newcommand{\UPos}{(0,0)}
\newcommand{\RMPos}{(7,0)}
\begin{table}[tb]
\centering
\begin{tabular}{|ll|}\hline
{\sf escrow $x$} & A request to escrow the credentials $x$,\\
& at the start of a journey.\\
{\sf finish $x$} & An attempt to identify to the receiving party,\\
& with parameter $x$, to end an existing journey.\\
{\sf finalise $x$} & A request to finalise the transaction of $x$, \\
&after being put into escrow.\\
{\sf rebate $x$} & A request for a rebate for unspent tokens, \\
& with parameter $x$.\\
\hline
\end{tabular}
\caption{Notation used in the subsequent figures depicting protocol extensions. This notation is an extension to the notation used in the original protocol specification found in the paper by Goodell and Aste~\cite{Goodell2019}.}
\label{table:lexicon}
\end{table}

The protocol has four separate stages as shown in Figure~\ref{fig: protocol_diagram}: Ticketing, Verification, System Entry, and System Exit.  In the Ticketing phase, the User initiates an interaction with the CP, akin to how a User might currently walk into a shop to buy an Oyster card.  The blind signature scheme is constructed in multiple stages, although it is the User device that performs all of the blinding and unblinding.

The difference between the adaptation shown in Figure~\ref{fig: protocol_diagram} and the original DigID protocol is that in the Ticketing stage, the Certification Providers (CP) send the signed (blinded) credential directly back to the user (step 4) in addition to publishing them to the ledger (step 5).  Later, the credential can be presented to an AP to authenticate with a Service.  The main reason that the CP must publish these blinded credentials to the ledger is so that users can verify the size and timing characteristics of the anonymity set.  Specifically, users expect that the CP does not treat specific users differently, for example by using different public keys for different users.  By allowing users to request an AP to send blocks from a requested interval to the user for inspection, the user can verify the timing and key usage of credentials issued by a CP.  The design of the original DigID protocol anticipates a use case wherein single-use credentials will be distributed to users over time via the ledger, to avoid unnecessary metadata leakage.  However, with the public transport use case, all of the single-use credentials (that is, tokens) can be immediately shared with users during the Ticketing interaction.  Although users must still have a way to verify the size and timing characteristics of the anonymity set, the tasks of verification and redemption can be separated in the public transport use case.  Figure~\ref{fig: scenario_solutions} illustrates this separation.  In the Verification stage, the user can anonymously request the associated proof block from an AP and use the information to verify both the size of the anonymity set and whether the tokens received from the CP match the ones published to the ledger.  However, because the user already has the tokens, the Verification stage is not a precondition to redeeming them.  The motivation for making the Verification stage optional is to ensure that users can request and verify the blocks associated with their credentials.  This can be done whenever they wish: before, during, or after using the transport system.  Users can then find and report any discrepancies.  Research by Heydt-Benjamin et al~\cite{privacy-for-public-transportation} shows that hybrid systems in which interested users can verify the claimed privacy assurances whilst data-restricted users are free to use the system may provide the similar security and anonymity benefits as systems where all users must perform verification.

\begin{figure*}[!ht]
  \centering
  \resizebox{0.99\textwidth}{!}{
  \begin{sequencediagram}
    \renewcommand\unitfactor{0.7}
    \newinst[0]{cp}{Certification Provider}{}
    \newinst[2.8]{user}{User}{}
    \newinst[2.8]{ap}{Authentication Provider}{}
    \newinst[2.8]{dlt}{Distributed Ledger}{}
    \newinst[2.8]{service}{Service}{}

    \begin{sdblock}{Ticketing}{}
    \path (0,0) -- (25,0);
        \begin{call}{user}{(1) User initiates request with a CP.}{cp}{(2) CP initiates blinding protocol.}
        \end{call}
        \addtocounter{seqlevel}{1}
        \begin{call}{user}{(3) User sends $[token]$. }{cp}{(4) CP signs $[token]$.}
        \end{call}
        \addtocounter{seqlevel}{1}
        \begin{call}{cp}{(5) CP publishes response from (4). }{dlt}{(6) Acknowledgement.}
        \end{call}
    \end{sdblock}    
    \begin{sdblock}{Verification}{}
    \path (0,0) -- (25,0);
        \begin{call}{user}{(7) User requests anonymity set.}{ap}{(8) AP sends anonymity set.}
        \end{call}
    \end{sdblock}    
    \begin{sdblock}{Usage - Entry}{}
    \path (0,0) -- (25,0);
        \begin{call}{user}{(9) User requests entry.}{service}{(10) Service sends nonce.}
        \end{call}
        
        \addtocounter{seqlevel}{1}
        
        \begin{call}{user}{(11) User presents signed tokens; sends $[nonce]$.}{ap}{(14) AP signs $[nonce]$.}
        \begin{call}{ap}{(12) AP adds tokens to escrow.}{dlt}{(13) Acknowledgement.}
        \end{call}
        \end{call}
        
        \addtocounter{seqlevel}{1}
        
        \begin{call}{user}{(15) User sends signed nonce, after unblinding.}{service}{(16) Service grants access.}
        \end{call}
    \end{sdblock}    
    \begin{sdblock}{Usage - Exit}{}
    \path (0,0) -- (25,0);
        \begin{call}{user}{(17) User requests exit by repeating (15).}{service}{(18) Service sends nonce.}
        \end{call}
        
        \addtocounter{seqlevel}{1}
        
        \begin{call}{user}{(19) User presents signed tokens; sends $[nonce]$.}{ap}{(22) AP signs $[nonce]$.}
        \begin{call}{ap}{(20) AP finalises token transaction.}{dlt}{(21) Acknowledgement.}
        \end{call}
        \end{call}
        
        \addtocounter{seqlevel}{1}
        
        \begin{call}{user}{(23) User sends signed nonce; requests rebate.}{service}{(24) Service grants exit and acts on rebate request.}
        \end{call}
    \end{sdblock}    
  \end{sequencediagram}
  }
  \caption{A schematic representation of the protocol. From the perspective of a passenger, each section would constitute a single interaction. Where $[x]$ is written, the value $x$ is blinded.}
  \label{fig: protocol_diagram}
\end{figure*}

Figure \ref{fig: scenario_solutions} shows the dual-stage approach that could be adopted to address the issue of calculating journey cost as described above, using message numbers from Figure~\ref{fig: protocol_diagram}.  Figure \ref{fig: scenario_solutions}$a$ shows the ``setup'' or \textit{entry} phase, wherein the user begins a journey. The AP uses the ledger to mark the tokens passed in the \textit{escrow} request as having been used; this request fails if the user has not proven ownership of the token.  Only after placing this token into escrow does the AP sign the nonce [$y$] (10, 14). In the second stage as shown by Figure~\ref{fig: scenario_solutions}$b$, the user repeats the request for a new nonce $z$ sent by the service. To identify the request, the user must in this stage supply the signature AP($y$), which it had received from the service in the entry phase (17). The assumption made here is that AP($y$) is a shared secret. The user then proves ownership of the credential used in the setup phase and requests a signature on [$z$] (19). The \textit{finalise} request should work similarly to the \textit{escrow} request: Whereas the \textit{escrow} request should mark a fresh token as `in escrow', the \textit{finalise} request should mark a token `in escrow' as having been `spent'.

Using AP($z$), the user then requests a rebate for the difference between the price of the journey and the amount of money required for escrow (23), the value of which could be calculated based on information provided in the nonces $y$ and $z$. There are a few alternative ways that a user might claim this rebate.  The simplest way that does not introduce additional de-anonymisation risk involves requiring the user to provides $[r]$, a set of blinded tokens corresponding to the value of the rebate, which the service can then send to the CP for signing, and the service can relay the blind signature back to the user, whilst the CP updates the ledger independently, as it does in the Ticketing stage.  However, it is important to consider that the system can allow the user to exit immediately rather than wait for the blind signatures.  The user has an incentive to collect the blind signatures and does no harm by failing to do so, so it is safe to allow the user to traverse the exit gate before fulfilling the rebate.  To maintain flow, the user could be given an opportunity to tap again, perhaps closer to the station exit, and indeed the user could complete the rebate operation later still, if desired.
\begin{figure*}[tb]
\begin{center}
\begin{center}
\scalebox{\figsize}{\begin{tikzpicture}[>=latex, node distance=3cm, font={\sf}, auto]\tss

\node (ap) at \APos [csm, draw, ultra thick, text width=4em] {\textbf{AP}};
\node (service) at \RMPos [csm, draw, ultra thick, text width=4em] {\textbf{Service}};
\node (user) at \UPos [csm, draw, ultra thick, text width=4em] {\textbf{User}};

\draw[->, line width=0.5mm] (user) edge[bend right=10] node[sloped,above,align=center] {
    (11) prove-owner $x^*$\\escrow $x$\\request CP($x$), [$y$]
} (ap);

\draw[->, line width=0.5mm] (ap) edge[bend right=10] node[sloped,below] {(14) AP([$y$])} (user);

\draw[->, line width=0.5mm] (user) edge[bend right=10] node[sloped,below] {(15) AP($y$)} (service);

\draw[->, line width=0.5mm] (service) edge[bend right=10] node[sloped,above] {(10) request $y$} (user);

\end{tikzpicture}}\\
\textit{{\bf\textit{(\ref{fig: scenario_solutions}a)}} A schematic representation of the System Entry stage.}
\end{center}
\vspace{5pt}
\begin{center}
\scalebox{\figsize}{\begin{tikzpicture}[>=latex, node distance=3cm, font={\sf \small}, auto]\tss

\node (ap) at \APos [csm, draw, ultra thick, text width=4em] {\textbf{AP}};
\node (service) at \RMPos [csm, draw, ultra thick, text width=4em] {\textbf{Service}};
\node (user) at \UPos [csm, draw, ultra thick, text width=4em] {\textbf{User}};

\draw[->, line width=0.5mm] (user) edge[bend right=10] node[sloped,above,align=center] {
    (19) prove-owner $x^*$\\
        finalise $x$, [$z$]\\
} (ap);

\draw[->, line width=0.5mm] (ap) edge[bend right=10] node[sloped,below] {(22) AP([$z$])} (user);

\draw[->, line width=0.5mm] (service) edge[bend right=5] node[sloped,below] {(18) request $z$} (user);

\draw[->, line width=0.5mm] (user) edge[bend left=20] node[sloped,above] {(17) finish AP($y$)} (service);

\draw[->, line width=0.5mm] (user) edge[bend right=20] node[sloped,below, align=center] {(23) AP($z$), rebate $r$} (service);
\end{tikzpicture}}\\
\textit{{\bf\textit{(\ref{fig: scenario_solutions}b)}} A schematic representation of the System Exit stage.}
\end{center}
\caption{A schematic representation of the interactions involving a passenger when entering or exiting the transport system.}
\label{fig: scenario_solutions}
\end{center}
\end{figure*}
In both extensions, $x$ is modelled as a single token. However, in practice, it may be necessary to use several tokens to obtain some value greater than that of a single token. This is possible by extending the meaning of $x$ to that of some token vector $\Bar{x}$, such that the requests work with several tokens. It is important to note that all operations should work on each element in the vector, not the vector itself.

Subsection \ref{privacy-idms} describes three main requirements of a privacy preserving IDMS: the minimisation of trust points, the inability to co-opt the system for mass surveillance, and the requirement for users to have control over their data linkages. Given the use of blind signatures where information is shared between parties it is clear that trust is not needed when the user interacts with the AP or the Service. A user might potentially need to trust a CP with their payment information, but this issue can be avoided by using cash to purchase tokens. The use of blind signatures reduces vectors that enable mass surveillance, because information is unlinkable between trips and between involved parties.  Finally, the protocol give users more control over their data linkages because the process of acquiring and spending tokens is directly controlled by the user without reference to an account. Users are able to generate their own tokens, and determine the size of the anonymity set they require.

\subsubsection{Technical Tradeoffs}
The discussion above has largely focused on functional requirements. It is, however, important to consider the various choices that users can make, which technical trade-offs they require, and how their choices might impact the feasiblity of meeting the functional requirements. Technical requirements are first discussed in a general context before the four specific use cases are discussed.

Table \ref{tab:technical-comparison} details how changes in the main network parameters affect the network performance metrics and provides a basis for the discussion of technical trade-offs. The number of system participants, such as node operators and end-users, increases network latency. Increasing the number of node operators increases network latency because, as opposed to consensus mechanisms, such as Proof of Work, wherein adding extra nodes increases the system's ability to achieve greater throughput\cite{Consensus-comparison}, a larger set of node operators requires more time to achieve consensus in many permissioned Byzantine Fault Tolerant-protocols~\cite{ScalableBFT}.  System architects, therefore, need to anticipate the expected number of transactions in a given time-frame (the throughput), as well as how quickly finality is required (the latency), to calculate performance under average and worst-case loads. Increasing the number of end users increases latency, assuming bandwidth remains equal because operators are limited by bandwidth and hardware when responding to user requests; therefore, increasing the number of users might lead to queued requests.

\begin{table}[!ht]
\centering
\begin{threeparttable}
    \begin{tabular}{@{}lcc@{}}
    \toprule
    \multicolumn{1}{c}{\multirow{2}{*}{Action}} &
    \multicolumn{2}{c}{Effect} \\ \cmidrule(l){2-3} 
    \multicolumn{1}{c}{}                        & Latency        & Throughput        \\ \midrule
    Increasing Network Nodes                               & \up{crossred}\tnote{1}  &      -         \\
    Increasing Network Users                               & \up{crossred}  &        -           \\
    More Security                                    & \up{crossred}  & \down{crossred}  \\
    More Precomputed Data                                    & \down{checkgreen}\tnote{*}  & \down{crossred}\\
    \bottomrule
    \end{tabular}
     \begin{tablenotes}
      \tiny
      \item[1.] When interactions involve writing to the distributed ledger.
    \end{tablenotes}
\end{threeparttable}
\caption{Tabular representation of the effect of increasing certain network parameters. The effect of decreasing the network parameters can be discerned by taking the opposite result presented. The symbol \UParrow\ indicates that the parameter increases the respective metric, whilst \DOWNarrow\ indicates a respective decrease. The symbol - indicates that the parameter does not affect the metric. The colours green and red should be interpreted as a favourable change and an unfavourable change respectively. Additionally a the symbol * has been added where the change is deemed favourable.}
\label{tab:technical-comparison}
\end{table}

System security is an important technical consideration because although DigID is designed to minimise trust between participants, they must trust the cryptographic primitives on which the protocol is built. These core primitives are blind signatures, digital signatures, and cryptographic hashes. The length of the key or hash is typically used as a proxy for the security of a given primitive. Therefore, the longer the key or digest the more secure a given primitive.  However, increased security comes at a price. Longer keys are typically slower to compute, resulting in requests taking longer to process and thereby increasing latency. Additionally, longer keys and hashes take up more space, thereby using up more bandwidth. This ultimately results in fewer requests communicated per second, thus reducing throughput. Therefore, there is a trade-off between security and latency, and the choice would depend on the type of transaction and the system's security assumptions.

The metadata transmitted alongside standard DigID protocol requests can play a large role in speeding up specific interactions. A specific example being storing (\textit{hash}, \textit{proof}) pairs instead of storing only the proofs in ledger blocks. This use of metadata allows the user to quickly index the block at the time that it is received instead of having to compute possibly thousands of hashes which would be unacceptable on a mobile device. Metadata, therefore, allows requests to be serviced faster, decreasing latency. Due to it using up more bandwidth, it can, however, decrease throughput.

\section{Implementation}\label{implementation}
To allow for fast and convenient development and testing, each system component that serviced requests was written in Python, using the Flask micro web framework~\cite{flask}. Each component was deployed in a \texttt{docker} container. In each container, a \texttt{gunicorn} server ran the component and requests were routed through an \texttt{nginx} proxy server.

Blind signatures are used in several instances throughout the protocol, and many of the privacy properties of the protocol rely on effective implementation of these signatures.\footnote{An alternative, valid approach to the design of this protocol involves zero-knowledge proofs; we do not explore that approach here.} One common blind signature scheme involves RSA signatures. The most common approach to implementing this scheme has inherent security issues related to the malleability of the ciphertext. Effort was made to find a reputable implementation suitable for our purposes that avoided these issues, although none was found.  We did not want to implement our own cryptographic library, so a Schnorr-based blind signature scheme proposed by Masayuki Abe~\cite{Abe} was used instead. The implementation of this scheme is based on the work of Antonio de la Piedra in the Charm Crypto library~\cite{blind_implementation}. The changes constituted refactoring it into classes and fixing type-related implementation bugs. Readers are referred to the original paper by Masayuki Abe~\cite{Abe} for proofs of correctness and a full protocol specification.

This blind signature scheme is used by token users when interacting with the CP to generate the requested credentials, and when interacting with the AP to blind the nonce generated by the service.  In the version of the protocol detailed in Figure~\ref{fig: protocol_diagram}, the challenge-response is both sent directly to the user (4) as well as published to the ledger alongside other proofs to make up the anonymity set (5).

 The only parameter that was needed to setup the blinding protocol was the security parameter that determines the group and hash size, which was set to $256$. For the use case of TfL, this was deemed sufficient. The security parameter was deemed sufficient for two reasons. First, analysis of Schnorr signatures, on which this scheme is based, suggests that for $b$ bits of security, the group must have size $3b$ and the hash must have size $2b$~\cite{Neven}\footnote{This proxy is imperfect because the schemes are not identical. Yet, it provides a valuable first estimate.}. Second, because there is a maximum monetary value associated with each block of proofs, if the cost of breaking a key exceeds the combined value of assets in the block, it can be argued that there is little motivation for doing so.  Therefore, although it is possible to brute force 85 bits of security, it is unlikely that this is economically practical for mass surveillance.

Finally, Hyperledger Fabric was used as the DLT in the implementation. Hyperledger Fabric was chosen because it can process a large number of transactions per second with low latency~\cite{Androulaki2018}. This mitigates the risk that the ledger becomes a bottleneck, even at peak times.

\section{Testing Methodology}\label{results}

To compare the performance of the implementation to that of existing production systems it is at least as important to understand why a system behaves the way it does, as it is to quantify its performance.  In the context of system optimisation, understanding why a system performs in a particular way allows developers the opportunity to mitigate unwanted behaviour.

With this requirement in mind, each API call was tested, the specifics of which are detailed below. These tests do not mimic the way a user would interact with a system; instead, they test the performance of the subroutines.  Each test is detailed below, along with the corresponding calls from Figure~\ref{fig: scenario_solutions}:\\

\begin{enumerate}
    \item \textit{Request nonce:} This call corresponds to messages (9) and (10) in Figure~\ref{fig: scenario_solutions}a, or messages (17) and (18) in Figure~\ref{fig: scenario_solutions}b.
    \item \textit{Prove ownership:} This call corresponds with the part of either messages (11) and (14) in Figure~\ref{fig: scenario_solutions}a or messages (19) and (22) in Figure~\ref{fig: scenario_solutions}b that includes the \textit{prove-owner} request, but does not include the interaction with the ledger for the \textit{escrow} or \textit{finalise} requests.
    \item \textit{Request signature:} This call corresponds with the part of either message (11) and (14) in Figure~\ref{fig: scenario_solutions}a or messages (19) and (22) in Figure~\ref{fig: scenario_solutions}b that includes the \textit{request} operation shown in request (2) of Figure~\ref{fig: scenario_solutions}a.  This call can be done in parallel with the \textit{Prove ownership} call.
    \item \textit{Verify block:} This API call corresponds with the part of either message (12) and (13) in Figure~\ref{fig: scenario_solutions}a or messages (20) and (21) in Figure~\ref{fig: scenario_solutions}b that correspond to the interaction with the ledger for the \textit{escrow} and \textit{finalise} operations.
    \item \textit{Verify signature:} This test corresponds to messages (15) and (16) in Figure~\ref{fig: scenario_solutions}a or messages (23) and (24) in Figure~\ref{fig: scenario_solutions}b that correspond to the verification of the signature required to permit the passenger to pass either the entry or exit gate, respectively.
\end{enumerate}

It should be noted that process for requesting credentials has not been included in the testing framework. The time required to process this request scales with the number of tokens requested. It is, however, possible to do much of the processing outside of the critical path, similarly to buying credit on the Oyster card website~\cite{Oyster_online}. The process of obtaining Oyster credit is much less time-bound than the use and validation at a fare gate, which is expected to require less than half of a second~\cite{MasterCard_2018}; latencies measured are therefore compared to this requirement.  Hypothesised bottlenecks are more relevant to the stages of the protocol in which the user enters and exits the system; for this reason, our tests focus on this those stages.

The implementation tested does not perfectly implement the protocol with extensions established for the TfL use-case in Section~\ref{overview}. Although the specific implementation details differ slightly, the implementation is an accurate proxy for performance. The two main differences are the fact that only the System Entry stage is tested, and that there is no explicit \textit{escrow} request. The current implementation is still an accurate proxy because the escrow command can be dealt with implicitly when proving ownership. No separate API calls are required, because the AP knows both the intent of the user and has all the primitives required to escrow the credentials.

We also assumed that the process of signing nonces would include blind signatures, as described in the protocol description.  From one perspective, blind signatures introduced at this stage are not particularly helpful to a user's privacy, since the service and the AP could collude to link a blinded nonce to an unblinded nonce.  However, we believe there is a data protection benefit from having the service and the AP not share the same identifier for a journey.  We also assume that where multiple signatures must be generated, the processing can be safely parallelised\footnote{This is not implemented in the proof of concept but is suggested for a production-level system.}, such that processing time depends only on the time taken to compute a single signature. Finally, we assumed that the call for the \textit{rebate} operation can be handled as described in Section~\ref{explanation}.  Therefore, we assert that the performance measured in the System Entry stage is a reasonable estimator for the performance of the System Exit stage.

The loads used during testing were one request per second and ten requests per second, which are described as the ``average'' load and ``maximum'' load, respectively. These loads are derived from the Rolling Origin and Destination Survey (RODS) collected by TfL~\cite{RODS-TfL}, which presented passenger arrival statistics for each station on the London Underground in 15-minute intervals. The assumption is made that arrivals within a 15-minute interval are uniformly distributed.  The reasons for this choice are that a uniform distribution is both easy to model and that additional information motivating another distribution is not available.  These loads are calculated as follows. First we define the set $\mathcal{I}$ as the set of all intervals at all stations $\mathcal{S}$. From this we obtain $\mathcal{I}_s \subset \mathcal{I}$, where $s\in\mathcal{S}$ such that $\mathcal{I}_s$ is the set of intervals of station $s$. An specific interval $i_t$ from this subset can then be obtained by selecting a specific time interval $t$, such that $i_t\in\mathcal{I}_s$. The average load is calculated by evaluating the following, and is the average of the maximum load across the different stations:
\begin{gather*}
    load_{avg} = \frac{\sum_{s\in\mathcal{S}}\max_t\mathcal{I}_s}{\vert \mathcal{S}\vert\times60\times 15}
\end{gather*}
The maximum load is calculated by evaluating the following:
\begin{gather*}
    load_{max} = \frac{\max_{s\in\mathcal{S}}\left(\max_t\mathcal{I}_s\right)}{60\times 15}
\end{gather*}

In both cases, the result was rounded up to the nearest integer. We divide by $15\times 60$ to transform load per 15 minutes to load per second. Finally, for all tests, loads were maintained for 20 minutes and run locally on a 3.3 GHz Dual-Core Intel Core i7-6567U CPU. Compared to current widely used mobile CPUs, such as the Exynos 8 Octa (8890) or Apple A10~\cite{Exynos_8890, A10_2020}, for which the i7-6567U is used as proxy, the i7-6567U is an inferior CPU. The i7-6567U has a lower CPU speed than both mobile CPUs as well as fewer cores that can be used for concurrent computation. We therefore believe that the results obtained form a conservative estimate for what is feasible.

The results presented aim to provide a sufficient description of the response time distributions for the test cases. They focus on the average cases for successful requests. MasterCard aims to limit the transaction time at ticket gates to 500 milliseconds and to achieve an average of 300 milliseconds~\cite{MasterCard_2018}. To compare our measured performance with the performance of the current system, we shall assume that 500 milliseconds for a payment entry is acceptable.
\begin{table*}[tb]
\centering
\begin{tabular}{@{}rrrrrrrrrr@{}}
\toprule
\multirow{2}{*}{Requests}              & \multirow{2}{*}{Min} & \multicolumn{5}{c}{Percentiles}  & \multirow{2}{*}{Max} & \multirow{2}{*}{$\sigma$}  & \multirow{2}{*}{$S_P$}        \\ \cmidrule(lr){3-7}
                                       &                      & 25 & 50 & 75 & 95 & 99           &                      &                            &                               \\ \midrule
\textit{Request nonce}             &          15          & 23 & 27 & 33 & 54 & 66               &        111           & 11                         & 0.82                          \\
\textit{Prove ownership}                 &          46          & 71 & 85 & 106 & 177 & 282        &        385           & 43                         & 0.84                          \\
\textit{Request signature}     &         119          & 289 & 394& \textbf{520}& \textbf{974}& \textbf{1308}&       \textbf{1742}   & 236                        & 0.70                          \\
\textit{Verify block}                     &           113         &193 &223 & 275& \textbf{635}& \textbf{779}       &        \textbf{953}  & 141                         & 0.98                         \\
\textit{Verify signature}  &           9         & 16 & 19 & 24 & 45 &  61                      &        111            & 10                          & 0.90                         \\ \bottomrule
\end{tabular}%
\caption{The response-time statistics (in milliseconds) of protocol requests under the average load experienced by the TfL network. The average load the TfL network experiences corresponds to 1 request per second. These loads were maintained for five minutes for each test. The distribution statistics are reported on successfully completed requests. Values in bold are those that fall above the 500 millisecond cut-off. $S_P$ denotes Pearson's median skewness.}
\label{tab:results_average_load}
\end{table*}

\begin{figure}[tb]
    \centering
    \begin{tikzpicture}
    \begin{axis}[
    title={Throughputs of High-Latency Requests},
    xlabel={Requests per Second},
    ylabel={Successful Requests Processed per Second},
    xmin=0, xmax=11,
    ymin=0, ymax=11,
    legend pos=north west,
    ymajorgrids=true,
    grid style=dashed,
    ]

    \addplot[color=gray, style=dashed, mark=None] table [x=x, y=control, forget plot] {data/incoming_vs_handled.dat};
    \addplot[color=red, mark=square] table [x=x, y=request] {data/incoming_vs_handled.dat};
    \addlegendentry{Request Signature}
    \addplot[color=black, mark=x] table [x=x, y=verify] {data/incoming_vs_handled.dat};
    \addlegendentry{Verify Block}
    \end{axis}
    \end{tikzpicture}
    \caption{The throughput of the two test cases for which the measured median latency was over 500 milliseconds in the maximum load scenario.}
    \label{fig: requests_vs_response}
\end{figure}
\begin{figure}[tb]
    \centering
    \begin{tikzpicture}
    \begin{axis}[
    title={Blind Signature Generation},
    xlabel={Integer Group Size [bits]},
    ylabel={Response Time [ms]},
    ymin=0, ymax=11,
    xtick={0,128,256,384,512,640,768,896,1024},
    ytick={0,2,4,8},
    yticklabels={2,4,8},
    ymode=log,
    log basis y={2},
    legend pos=north west,
    ymajorgrids=true,
    grid style=dashed,
    ]

    \addplot[color=blue, mark=triangle] table [x=x, y=mean] {data/blindsign.dat};
    \end{axis}
    \end{tikzpicture}
    \caption{Time taken to generate a single blind signature.}
    \label{fig:blind_sign_vs_group_size}
\end{figure}
\begin{table}[tb]
\centering
\begin{tabular}{@{}lr@{}}
\toprule
Request & Response Size (kB)\\ \midrule
\textit{Request Nonce} & 0.67 \\
\textit{Prove Owner} &  1.79\\
\textit{Request Signature} & 5.20 \\
\textit{Verify Signature} &  0.15\\ \midrule
\textit{Total} &  7.81\\ \bottomrule
\end{tabular}%
\caption{The response sizes in kilobytes of the sub-requests made in Access Service. The Access Service request in its simplest form does not contain the Verify Block request and is therefore not shown. The table does not include data sent in the request.}
\label{tab:response_sizes}
\end{table}
\section{Results}
\label{analysis}
This section presents the results obtained in the experiments and explores the patterns observed.  Possible solutions are suggested for all the shortcomings discussed.  The main takeaway is that a privacy-enhancing digital payment system for public transport is feasible, and that we describe a robust methodology for improving its implementation.  Although total request latency in our test environment exceeded 500 milliseconds, the bottlenecks are known, and potential solutions are both known and feasible to implement. We also briefly analyse the possible economic costs of the proposed solution.

\subsection{Performance Analysis}

One of the research questions was to what extent a payment system based on DigID can potentially meet or even exceed the performance requirements of Transport for London's existing payment infrastructure. Therefore, we compare results to what is currently being achieved by the TfL system and suggest possible improvements where our implementation is currently insufficient.

First, we consider the tests of the implementation of our protocol in terms of network requirements and in terms of latency.  Table~\ref{tab:response_sizes} depicts the response sizes of the individual sub-requests.  These values were collected alongside the load data and therefore do not include the outgoing request sizes.  In general, the responses are larger than the request sizes.  In the TfL use-case, multiple public keys might need to be escrowed simultaneously, which would result in a linear increase in request and response sizes. The amount of credit associated with each key and the corresponding ease of use will, therefore, need to be traded off against data usage.

At high levels of concurrent users, fewer requests are processed than received per second, creating a queue, which increases latency.  Figure~\ref{fig: requests_vs_response} shows how the system responds.  The figure shows the difference in requests made and requests successfully processed under varying loads. The dashed diagonal line shows the point at which the number of incoming requests equals the number of processed requests. Results under this diagonal indicate queue formation.

Two approaches are suggested to mitigate this issue and reduce the latency to acceptable levels as experienced at lower loads. First, the number of parallel workers available can be increased, such that the load can be balanced between them. Second, the protocol can be further optimised to reduce request processing time. In this case, a single worker will be able to process requests in less time, such that maximum throughput is increased. Mathematically re-implementing heavy tight loops in C/C++ or moving the blinding protocol onto an FPGA could lead to an improvement that would render performance sufficient for the needs of TfL. Fast implementations of cryptographic algorithms, such as error-correcting codes, are widely available~\cite{Chelton2008} and their use is considered common practice in the field of cryptocurrency mining.

Figure~\ref{fig: requests_vs_response} shows that, given the current configuration, users of our implementation will start experiencing unacceptable queueing if there are more than five concurrent users per second. According to the RODS dataset~\cite{RODS-TfL}, for all stations between 5 am and midnight this load only occurs 0.63\% of the time, and all occurrences are contained to nine stations in the TfL network.  Each of these stations has numerous ticket gates. Groups of these ticket gates could be aggregated to run on separate hardware and be treated as `sub-stations'. The correct configuration would decrease the number of concurrent users on each system and reduce load to acceptable levels.

The System Entry stage of the protocol, which is used by passengers when starting a journey, comprises five subroutines: \textit{request nonce}, \textit{prove ownership}, \textit{request signature}, \textit{verify block}, and \textit{verify signature}.  The \textit{prove ownership} and \textit{request signature} calls can run in parallel, which is to say that the time required for both of the calls can be limited to the time required for the longer of the two.  With this assumption, estimates of the total amount of time required for the entire interaction at the 50th (median) and 95th percentiles, assuming that performance of the different subroutines are perfectly correlated, are 663 milliseconds and 1.708 seconds, respectively.

There are two possible improvements that we can make to reduce the time required for the interaction.  The first improvement is to accept that although blind signature generation at the Ticketing stage is necessary to protect the privacy of a passenger, blind signature generation is actually not necessary in the System Entry and System Exit stages and can be replaced with ordinary signature generation instead.  The second improvement is to move the DLT operations outside the critical path and rely upon a separate mechanism to ensure operational security.  We consider both of these approaches in turn.

\subsubsection{Simplifying signature requests upon entry and exit}

As described earlier in this section, the purpose of using blind signatures in the \textit{request signature} subroutine of the System Entry and System Exit stages is to avoid sharing data about individual transactions between the Service and the AP.  This has the benefit of simplifying compliance with data protection rules and protecting against offline attacks performed \textit{ex post} without transaction records.  However, from the perspective of a passenger, telling the AP and the service about a particular trip is not harmful, provided that there is no harm in the AP or services learning information about both endpoints of the journey itself.  Assuming that the passenger is primarily concerned about linking the journey to his or her identity, the blind signature operation performed in the Ticketing stage is sufficient.  Additionally, blind signatures at the System Entry and System Exit stages do not protect against timing attacks performed either in real time or with the aid of transaction logs.  There is no way for a passenger to verify that such an attack has not been carried out, so it can be argued that the privacy benefit of using blind signatures in this stages is small.

Of all the subroutines, the \textit{request signature} subroutine is the largest contributor to the latency experienced by passengers.  It was hypothesised that the blind signing process contributed heavily to this latency. Figure~\ref{fig:blind_sign_vs_group_size} shows the result of an experiment to identify how the execution time of a blind signing process varies depending on the size of the integer group in which the cryptographic arithmetic takes place. This experiment ignores networking overhead and focuses on the execution speed of the cryptographic functions. The range of values on the independent axis was deemed sufficient to characterise both the relationship shown and to depict the signing times for probable choices in group size ranging from low-security (128 bits) to high-security (1024 bits).  The hypothesis that the signing functions contribute heavily to the \textit{request signature} latency was not substantiated by the evidence. The minimum response time of \textit{request signature} was 112 milliseconds while the execution time of the blind signing process takes 1.31 milliseconds with a group size of 256 bits. The slow response time is likely due to the combination of two factors. First, \textit{request signature} has comparatively larger request bodies that need to be marshalled. Second, there are additional network requests that are needed in the setup stages of the blind signature protocol. We argue that the latter explanation is the dominant factor. Here, the only variable is the security parameter of the blinding protocol. A higher security parameter leads to larger signatures. Although the request bodies can be optimised further by removing redundancy\footnote{The current implementation encodes a value together with its modulus. The protocol that is being used is known \textit{a priori}, and these moduli are defined only once. Removing these moduli from subsequent requests would reduce the size of the request body and could speed up marshalling and un-marshalling.}, it is unlikely that much can be done to reduce the fixed networking costs.

Because removing the blinding operations from the \textit{request signature} subroutine would reduce the number of network interactions and cryptographic operations to the same used in \textit{prove ownership}, and that the \textit{prove ownership} and \textit{request signature} operations can be parallelised, we can potentially improve the speed by the difference in latency between the \textit{request signature} and \textit{prove ownership} subroutines.  The result is that at the 50th (median) and 95th percentiles, assuming perfect correlation between performance of the subroutines, latency experienced by the passenger would be 354 milliseconds and 911 milliseconds, respectively.  This performance would be acceptable in the average case.

\subsubsection{Removing DLT interactions from the critical path}

The other possible improvement to the protocol is to remove interaction with the distributed ledger from the critical path.  Although the use of a distributed ledger is central to the design of this architecture, removing it from the System Entry and System Exit stages might not be as far-fetched as it seems.

In this case, we might allow the process of committing the \textit{escrow} to the DLT will take place after access to the system is granted, and optimistically assume that it will successfully finish in a short period of time (say, a minute) after the fare gates grant access to the service, that the prospective passenger will take at least that long to either tap out from an exit gate or attempt to cancel the ride, and that any double-spending would be identified either by exit gates or by ticket inspectors.  In this case, we would need a separate mechanism to ensure that exit gates or ticket inspectors could verify escrow state without a DLT operation, such as (a) linking the escrow (on the distributed ledger) to a specific AP or other service provider that must be contacted in the System Exit stage to verify the exit, or (b) requiring riders to carry trusted hardware that would link a unique device to the unique entry nonce.  Of course, the decision to circumvent the call to the DLT during the System Entry and System Exit stages carries costs as well, either by introducing some measure of fragility, both by assuming that the DLT transaction will complete before the user seeks to exit the system and by binding the System Entry to a particular central actor, or by requiring the user to carry a device within the security envelope of a system operator, which introduces many problems~\cite{goodell2022}.

Nevertheless, if the system were designed with these trade-offs in mind, then the \textit{verify block} subroutine could be removed from consideration when calculating latency.  Under this assumption, at the 50th (median) and 95th percentile, assuming perfect correlation between performance of the subroutines, latency experienced by the passenger would be 440 milliseconds and 1.073 seconds, respectively, which would be acceptable in the average case.  If this improvement were combined with removing the blinding operations from the System Entry and System Exit stages, then latency would be reduced further still, potentially to 131 milliseconds and 276 milliseconds, respectively, which would be acceptable in nearly all cases.  We conclude that subject to some trade-offs, it is feasible to implement a privacy-respecting, decentralised payment system with user interactions that are similar to those used today by TfL.

\subsection{Cost Analysis}

TfL supports payment both via the Oyster system, which it oversees, as well as through contactless bank card and mobile phones via mechanisms that are managed externally.  To avoid introducing significantly new workflows for passengers, it should be possible to be able to use our proposed protocol both from a mobile phone as well as from a card or smart device that is similar to the Oyster cards currently issued by TfL.

The case for feasibility on mobile phones has already been briefly explored in this section.  Modern mobile phones have sufficient storage, computation, and communication capabilities that users should expect no marginal costs to use the new protocol beyond the purchase of credit to use the system. We therefore focus on the cost feasibility of a smart device issued by TfL.

Estimating the exact cost of manufacturing is outside the scope of this work. Instead, as a proxy for feasibility, we use the costs of the components we expect to be vital for such a device. We anticipate that the main components of such a device would include a radio frequency (RF) transceiver, a micro-controller, an FPGA, a battery, and a container. If we assume that these items are purchased in bulk, say thousands of units per order, by the issuing authority, we observe that the total unit price is under £10, without including wiring and manufacturing costs\footnote{RF Transceiver: £1.53 per unit~\cite{SX1233IMLTRT}. Micro-controller: £5.70 per unit~\cite{R5F572TKCDFB}. FPGA: £0.892 per unit~\cite{ICE40LP384-SG32}. Battery: £0.473 per unit~\cite{SR626}. Container: £0.854 per unit~\cite{PB-1574}}.
\section{Conclusions}\label{conclusion}
This chapter explores how, given a specific use case, a privacy-enhancing, decentralised token system compares to the TfL payment systems in public transport with respect to its design requirements. Furthermore, we limit the scope of the work to the following three main research questions:

\begin{enumerate}
    \item How does the DigID address the key privacy and security requirements of modern public transport payment systems compared to current centralized solutions?
    
    \item What are the technical, operational, and user experience trade-offs when implementing DigID in a large-scale public transport network?
    
    \item To what extent can DigID meet or exceed the performance requirements of Transport for London's existing payment infrastructure?
\end{enumerate}

First, if implemented correctly, this protocol would allow a user to travel while interacting with the transport system in a manner nearly identical to what can be experienced when using the Oyster card. Yet, unlike when using the Oyster network, none of the trips taken by a given user can be linked to each other. This offers greater privacy, because users cannot be profiled based on their travel habits or patterns. Additionally, because payment information is kept separate from travel habits, travel history can remain private even if the user uses non-private means of payment, such as a bank card, to acquire tokens.

We note that although there are several technical parameters to consider, such as the number of network nodes, key lengths and other security parameters, and the format of data on the ledger, these have been de-risked in terms of their influence on the user experience because most of their impact is upon processes that can run in the background and do not affect the ``hot loop'' of real-time interaction.  Nevertheless, they remain key design choices for a system implementer.

It may also be concluded that a workable solution is possible using common hardware. At a service level, the performance of the protocol is unlikely to be hardware-constrained and should be able to comfortably meet demand in the TfL use case. The identified bottleneck of signature generation can be traded off without sacrificing anonymity for passengers.  Finally, given the memory requirements of the protocol, it is possible to conclude that the protocol can be run on mobile devices. Currently, it is not possible to say whether the data usage would be reasonable because it depends on variable factors that need to be set by the implementer such as token value and a more detailed analysis of what could be considered ``reasonable usage.'' Given the variable parameters, however, the usage could be calculated in advance and left to the discretion of the user.

Addressing the final sub-question, the results show that for a throughput of one user per second, it is possible to process this load in under 500 milliseconds with off-the-shelf hardware, given some adaptations.  In the average case, our proposed protocol can compete with existing solutions. However, to achieve this result in worst-case situations requires additional improvement.  More generally, it can be concluded that the proof-of-concept demonstrates that, given appropriate hardware, a system following our proposed protocol can be used for fare-collection with performance comparable to the current fare-collection system.

\subsection{Evaluation}\label{evaluation}

The experimental setup is viewed as a useful `first step' in discerning the viability of the protocol for the TfL use case. Yet, there are some important limitations to highlight.

First, we assume a security parameter of $n=256$ is sufficient for the blinding protocol. This a compromise between network latency and security. Although it is possible to un-blind the credentials by brute force, it isn't deemed useful economically. Ideally, the security parameter would be increased to $n=512$ or more, to remove to the possibility of advanced persistent threats involving the de-anonymisation of credentials, but it might be necessary to combine this with the various optimisations that we described.

Additionally, all the experiments were run locally, on a single device. This can be regarded as a strength because it provides a more resource-constrained environment compared to deployment on a separate server. Therefore a protocol that can meet the requirements in this environment, is more likely to work well in production. It does, however, have a few weaknesses. First, because everything is run locally, the results do not take network latency into account. It is not possible to comment on the extent to which this affects feasibility.  However, we assume that a local network can serve as a proxy for Near Field Connections. Second, this protocol will be run on phones or other smart devices. Although we substantiate the argument that the current results form a useful proxy, this is still an inference. Both of these limitations can be tested, and their impact assessed, with further investigation.

Certain parts of the proposed extensions were not implemented. These extensions include the System Exit stage of the protocol.  Although we were able to provide an estimate of the performance of both the System Entry and System Exit stages, our measurements might not capture specific implementation details that significantly affect performance.  We also note that we assumed that factors affecting the performance of different subroutines were correlated.  We argue that this assumption is conservative, but it might not capture interaction effects that we had not considered.

\subsection{Future Work}

Although this project offers a partial answer the question of viability of a privacy-respecting, decentralised payment system for public transport in a production setting, there are important shortcomings and areas that should be further explored before it can be deployed into production. We offer some reasoning for our prioritisation of such work.

\begin{enumerate}
    \item We note that the experiments were run locally, which may affect system performance, and subsequent conclusions are drawn. It is recommended that future work start by redeploying the current system on infrastructure that more accurately reflect a production environment. We recommend redeploying the system on multiple nodes in order to simulate CPs, APs, and Services, as well as running user code on various mobile devices. 
    \item Extend the implementation such that it implements the complete System Entry and System Exit subroutines. It was mentioned that the proof-of-concept does not perfectly implement the protocol description specified in Section~\ref{overview}, by first addressing the deployment issue (1), the results from this work will be more accurate.
    \item Next, the extent to which the messages can be optimised should be investigated. Doing so would permit the analysis of a system that more closely resembles the production system and is more likely to meet TfL requirements in all cases. Effective optimisation requires an understanding of the merits and weaknesses of a system, otherwise efforts might be misplaced. This is why we recommend first exploring the above two areas.
    \item Finally, it cannot be expected that all users have equal access to the Internet, nor can Internet access be guaranteed in underground stations. Our protocol can be run as long as infrastructure similar to that provided by Cubic is internally networked and can communicate with user devices locally. Work can be done to adapt the implementation to enable offline or near-field use and analyse effects on performance.
\end{enumerate}

All of these recommendations are considered actionable and would further prove the robustness of the protocol.
\bibliographystyle{apalike}
\bibliography{bibliography}

\end{document}